\documentclass[twocolumn]{aastex631}

\usepackage{subfigure, graphicx, amsmath}

\newcommand{\hst}{\textit{HST}}
\newcommand{\jwst}{\textit{JWST}}

\newcommand{\um}{$\mu$m}

\newcommand{\ah}{$A_{H}$}

\newcommand{\water}{$\mathrm{H}_2\mathrm{O}$}
\newcommand{\methane}{$\mathrm{CH}_4$}

\submitjournal{ApJL}

\begin{document}

\title{Clouds and Clarity: Revisiting Atmospheric Feature Trends in Neptune-size Exoplanets}
\shorttitle{Clouds and Clarity}
\shortauthors{Brande et al.}

\author[0000-0002-2072-6541]{Jonathan Brande}
\affiliation{Department of Physics and Astronomy, University of Kansas, 1082 Malott, 1251 Wescoe Hall Dr., Lawrence, KS 66045, USA}
\correspondingauthor{Jonathan Brande}
\email{jbrande@ku.edu}

\author{Ian J. M. Crossfield}
\affiliation{Department of Physics and Astronomy, University of Kansas, 1082 Malott, 1251 Wescoe Hall Dr., Lawrence, KS 66045, USA}

\author[0000-0003-0514-1147]{Laura Kreidberg}
\affiliation{Max Planck Institute for Astronomy, K\"{o}nigstuhl 17, D-69117 Heidelberg, Germany}

\author[0000-0002-4404-0456]{Caroline V. Morley       }
\affiliation{Department of Astronomy, University of Texas at Austin, Austin, TX, USA}

\author{Travis Barman         }
\affiliation{Lunar and Planetary Laboratory, University of Arizona, Tucson, AZ 85721 USA}

\author{Bj\"{o}rn Benneke        }
\affiliation{Departement de Physique, and Institute for Research on Exoplanets, Universite de Montreal, Montreal, Canada}

\author[0000-0002-8035-4778]{Jessie L. Christiansen}
\affiliation{Caltech/IPAC-NASA Exoplanet Science Institute, Pasadena, CA 91125, USA}

\author[0000-0003-2313-467X]{Diana Dragomir        }
\affiliation{Department of Physics and Astronomy, University of New Mexico, Albuquerque, NM, USA}

\author[0000-0002-9843-4354]{Jonathan J. Fortney      }
\affiliation{Department of Astronomy and Astrophysics, University of California, Santa Cruz, CA 95064, USA}

\author[0000-0002-8963-8056]{Thomas P.\ Greene            }
\affiliation{NASA Ames Research Center Space Science and Astrobiology Division M.S. 245-6 Moffett Field, CA 94035, USA}

\author[0000-0003-3702-0382]{Kevin K. Hardegree-Ullman}
\affiliation{Steward Observatory, The University of Arizona, Tucson, AZ 85721, USA}

\author[0000-0001-8638-0320]{Andrew W. Howard         }
\affiliation{Cahill Center for Astronomy \& Astrophysics, California Institute of Technology, Pasadena, CA 91125, USA}

\author[0000-0002-5375-4725]{Heather A. Knutson       }
\affiliation{Division of Geological and Planetary Sciences, California Institute of Technology, Pasadena, CA 91125, USA}

\author[0000-0003-3667-8633]{Joshua D. Lothringer     }
\affiliation{Physics Department, Utah Valley University, 800 West University Parkway, Orem, UT 85058-5999, USA}

\author[0000-0001-5442-1300]{Thomas Mikal-Evans}
\affiliation{Max Planck Institute for Astronomy, K\"{o}nigstuhl 17, D-69117 Heidelberg, Germany}

\begin{abstract}

Over the last decade, precise exoplanet transmission spectroscopy has revealed the atmospheres of dozens of exoplanets, driven largely by observatories like the \textit{Hubble} Space Telescope. One major discovery has been the ubiquity of atmospheric aerosols, often blocking access to exoplanet chemical inventories. Tentative trends have been identified, showing that the clarity of planetary atmospheres may depend on equilibrium temperature. Previous work has often grouped dissimilar planets together in order to increase the statistical power of any trends, but it remains unclear from observed transmission spectra whether these planets exhibit the same atmospheric physics and chemistry. We present a re-analysis of a smaller, more physically similar sample of 15 exo-Neptune transmission spectra across a wide range of temperatures (200-1000 K). Using condensation cloud and hydrocarbon haze models, we find that the exo-Neptune population is best described by low cloud sedimentation efficiency ($\mathrm{f_{sed}}\sim0.1$) and high metallicity ($100\times$ Solar). There is an intrinsic scatter of $\sim0.5$ scale height, perhaps evidence of stochasticity in these planets’ formation processes. Observers should expect significant attenuation in transmission spectra of Neptune-size exoplanets, up to 6 scale heights for equilibrium temperatures between 500 and 800~K. With \jwst{}’s greater wavelength sensitivity, colder ($<500$~K) planets should be high-priority targets given their clearer atmospheres, and the need to distinguish between the ``super-puffs’’ and more typical gas-dominated planets.

\end{abstract}

\keywords{Exoplanet atmospheres (487) --- Exoplanet atmospheric composition (2021) --- Exoplanet atmospheric dynamics (2307) --- Transmission spectroscopy (2133) ---  Hubble Space Telescope (761) --- James Webb Space Telescope (2291)}

\section{Introduction} \label{sec:intro}

Detailed characterization has long been the goal of exoplanet astrophysics, but only in recent years has this truly been possible. In most cases, exoplanet detection studies have limited ability to probe conditions on these planets directly, leaving exoplanet characterization as a more targeted and expensive second-echelon effort. Characterization efforts can take different forms, but among the most successful has been the atmospheric characterization of gaseous exoplanets, through transmission, emission, or direct spectroscopy. Currently, direct spectroscopy requires sufficiently high-contrast imaging of widely separated, self-luminous exoplanets, while transmission and emission spectroscopy require these planets to transit their host stars, but also require sufficiently warm and large planets to produce detectable atmospheric absorption or planet-star contrast. Given current observational capabilities, transmission spectroscopy has been one of the standout successes in atmospheric characterization, driven primarily by ground-based high-resolution spectrographs as well as space-based observatories like the workhorse Hubble Space Telescope (\hst{}). With over 5500 exoplanets known, of which $\sim 4100$ transit \citep{ExoArchive}, we now have hundreds of planets characterized through transmission spectroscopy or spectrophotometry. 

However, these successes are not without their limits. These instruments often have limited spectral ranges, necessitating the use of multiple observatories, filters, and observations to observe multiple transits of an exoplanet in order to assemble broad-band atmospheric transmission spectra, and many well-known transit spectra are markedly flat \citep[e.g., GJ~1214~b,][]{Kreidberg2014}, either due to opaque cloud layers, atmospheric hazes, or the muting of absorption features by high atmospheric metallicity, limiting our ability to determine the content of these atmospheres. A major exception has been the use of \hst{}/WFC3 to observe water vapor absorption in dozens of exoplanetary atmospheres. \water{} has a well-known absorption feature near 1.4 \um{}, right in the middle of \hst{}/WFC3's G141 grism. In clear atmospheres, this feature can be quite prominent, allowing for the direct comparison of the cloudiness or clarity of observed exoplanet transmission spectra. Where high-altitude clouds or hazes are present, observed spectra show either no water absorption or muted features, while clear atmospheres show strong water absorption. 

Previous work \citep{Stevenson2016, Crossfield2017, Fu2017, Gao2020, Yu2021, Dymont2022, Edwards2023, Estrela2022} has attempted to find relationships between physical exoplanetary parameters (e.g. planet equilibrium temperature, $T_{eq}$ or surface gravity $g$) and the strength of observed atmospheric features in transmission. \cite{Stevenson2016} performed the first systematic study of this, and introduced the method of examining clarity through the WFC3 water feature amplitude \ah{} (although focusing on a surface gravity-selected sample), while initial results from \citep{Crossfield2017} found a linear trend in \ah{} for a small sample of 6 warm Neptunes ($R_p \in [2,6]$~R$_\oplus$, $T_{eq}\in [500, 1100]$~K), and showed that hotter planets tend to have less aerosol obscuration than cooler planets. \cite{Fu2017} found a similar positive linear trend for a larger sample which also included transmission spectra from hotter, more massive planets, and proposed a physical explanation based on cloud condensation instead of the haze formation hypothesis from \cite{Crossfield2017}. \cite{Gao2020} focused on hot (800~K~--~2600~K) giant exoplanets, measuring the water feature amplitude compared to model predictions, and found that all but the hottest planets in their sample must be attenuated by clouds. Higher-order polynomial trends have been a popular model for this behavior. \cite{Yu2021} fit a parabolic trend to a sample of 9 Neptune-sized planets, but were primarily conducting an analysis of haze evolution and did not elaborate on the implications of this trend. \cite{Dymont2022} examined a larger sample of 25 planets between 200~K and 1000~K, but included rocky planets as small as $1.13~R_\oplus$ (GJ~1132~b), and Jovian planets as large as $12.9~R_\oplus$ (WASP-67~b), and also found hints of the quadratic trend from \cite{Yu2021} (without fitting any models) between $T_{eq}$ and \ah{}, which depends entirely on K2-18~b and LHS~1140~b. Taking a Neptune-sized 16 planet subset of their full 70-planet analysis, \cite{Edwards2023} find a parabolic trend matching theoretical predictions from \cite{Kawashima2019b}, while \cite{Estrela2022} fit second- and fourth-order polynomials to 62 planets from 500 to 2500 K, finding two distinct populations: aerosol-free and partially cloudy/hazy planets, with minimal aerosol presence near 1500 K.

Here, we present a trend analysis of a homogeneous sample of Neptune-like exoplanets, in order to show that sample selection is important in order to identify atmospheric feature trends, especially as \jwst{} operations continue and provide higher-quality, broader-wavelength characterization of planetary atmospheres than possible from \hst{} alone.

\section{Sample Selection}
\subsection{Atmospheric Motivation}
Typically, observed features are characterized by atmospheric scale heights, converting from relative units (planet-to-star radius ratio or transit depth) to physical units (kilometers), where the atmospheric scale height of a planet is \citep{Deeg2018}: 
\begin{equation}
    H = \frac{k_B T_{eq}}{\mu g}
\label{eq:H}
\end{equation}  and the depth of an absorption feature at a particular wavelength, measured in numbers of scale heights, is: 
\begin{equation}
    \delta_\lambda = \frac{(R_p + nH)^2}{R_s^2} - \frac{R_p^2}{R_s^2} \approx \frac{2nR_pH}{R_s^2}
\label{eq:delta}
\end{equation}

Here, $K_B$ is Boltzmann's constant, $T_{eq}$ the equilibrium temperature of the planet, $\mu$ the atmospheric mean molecular weight, $g$ the planet's surface gravity, $R_p$ the planet radius, $R_s$ the stellar radius, and $n$ the number of scale heights.

However, this implies that the strength of an observed absorption feature may actually differ significantly depending on the assumed atmospheric metallicity. For example, two identical planets, at $1\times$ Solar metallicity ($\mu~\approx~2.3$~amu) and $100\times$ Solar metallicity ($\mu~\approx~3.05$~amu) will have $H_{100}/H_{1}~\approx~0.75$, and so identical-depth absorption features in a $100\times$ Solar metallicity atmosphere will be ``stronger'', i.e., correspond to more scale heights than a Solar metallicity atmosphere ($n_{100} \approx 1.33 n_1$). 

Since atmospheric metallicity appears to vary across the exoplanetary mass range \citep{Welbanks2019} and is expected from the solar system and from planetary formation models \citep{Fortney2013}, we must look at physically similar planets in a relatively narrow mass range to mitigate any mass-metallicity effect in order to have any hope of comparing their observed atmospheric features and finding underlying trends. We chose 15 planets for our analysis, including those from \cite{Crossfield2017} and others which have subsequently been observed with \hst{} WFC3. We started with a specific radius range ($\mathrm{R_\mathrm{pl}}\in[2,6]$~$\mathrm{R_\oplus}$), in order to ensure we were not including rocky super-Earths, and included a few larger planets (HAT-P-26 b, due to it's Neptune-like mass, and Kepler 51 b and d as, despite their low masses, they have well-known extended atmospheres and relatively low equilibrium temperatures). We also restricted our sample to temperatures $<1000$ K, in order to make sure we probe thermally-consistent atmospheric physics. As such, we do not include planets like 55 Cnc e, as it is nearly 1000 K hotter than any other planet in our sample, likely rocky, with no extended atmosphere, and has been shown not to have water vapor in its atmosphere. These planets and their physical parameters are shown in Table \ref{tab:sample}.

\begin{figure}[h]
    \centering
    \includegraphics[width=0.45\textwidth]{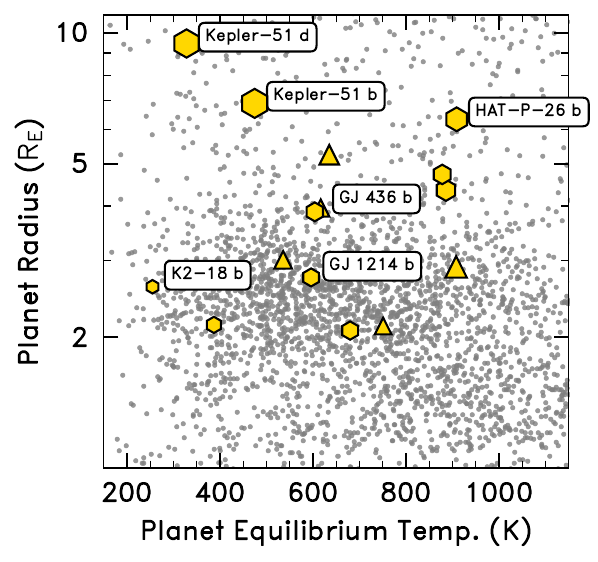}
    \caption{This work's sample overlaid with known transiting planets. Hexagons indicate planets in our sample also being observed by \jwst{} through Cycle 2, while triangles are not yet scheduled or approved for \jwst{} observations. Selected targets have been labeled.}
    \label{fig:my_label}
\end{figure}

\begin{deluxetable*}{lrrrrcrrrrrl}
    \rotate
    \tablehead{\colhead{Planet (Spectrum Ref.)} & \colhead{R$_{\mathrm{pl}}$} & \colhead{M$_{\mathrm{pl}}$} & \colhead{g$_{\mathrm{pl}}$\tablenotemark{*}} & \colhead{P$_{\mathrm{orb}}$} & \colhead{R$_*$} & \colhead{M$_*$} & \colhead{T$_{\mathrm{eff}}$} & \colhead{T$_{\mathrm{eq}}$\tablenotemark{*}} & \colhead{H\tablenotemark{*}} & \colhead{\ah{}\tablenotemark{*}} & \colhead{Parameter Ref.}\\
     & \colhead{$\mathrm{R_{\oplus}}$} & \colhead{$\mathrm{M_{\oplus}}$} & \colhead{$\mathrm{cm\,s^{-2}}$} & \colhead{days} & \colhead{$\mathrm{R_\odot}$} & \colhead{$\mathrm{M_\odot}$} & \colhead{$\mathrm{K}$} & \colhead{$\mathrm{K}$} & \colhead{$\mathrm{km}$}} 
    \startdata
    K2-18 b \citep{Benneke2019a}            &   2.61    &   8.63 	&   1241   	&  32.9400 	& 0.44 	&  	0.50 & 3457	& 253 &  56 	& 	$3.23_{-0.88}^{+0.98}$  &   \cite{Benneke2019a} \\
    Kepler-51 d \citep{Libby-Roberts2020}   &   9.46    &   5.7   	&   62 		& 130.1845  & 0.88  & 	0.98 & 5670 & 332 &  1450 	&  	$0.13_{-0.08}^{+0.24}$  &  	\cite{Libby-Roberts2020} \\
    TOI-270 d \citep{Mikal-Evans2023}       &   2.13    &   4.78 	&   1029  	&  11.3796  & 0.38 	&   0.39 & 3506 & 354 &  94     &   $3.50_{-0.91}^{+1.06}$  &  	\cite{VanEylen2021} \\
    Kepler-51 b \citep{Libby-Roberts2020}   &   6.89    &   3.69 	&   76    	&  45.1542  & 0.88  &   0.98 & 5670 & 472 &  1691   &   $0.34_{-0.25}^{+0.41}$  &   \cite{Libby-Roberts2020} \\
    HD 3167 c \citep{Mikal-Evans2021}       &   3.01    &   9.8  	&   1059   	&  29.8454 	& 0.86  &   0.86 & 5261 & 508 &  131    & 	$1.76_{-0.89}^{+0.90}$  &   \cite{Christiansen2017} \\
    GJ 1214 b \citep{Kreidberg2014}         &   2.74    &   8.17   	&   1064	&   1.5804  & 0.21  &   0.18 & 3250 & 536 &  137  	& 	$0.09_{-0.03}^{+0.07}$  &  	\cite{Cloutier2021} \\
    GJ 3470 b \citep{Benneke2019b}          &   3.88    &   13.73   &  	893 	&   3.3366  & 0.48 	&   0.51 & 3652 & 597 &  182   	& 	$1.03_{-0.26}^{+0.20}$  & 	\cite{Biddle2014} \\
    GJ 436 b  \citep{Knutson2014}           &   3.96    &   21.7\tablenotemark{$\dag$}   &   1355    &   2.6439  & 0.44  &   0.45 & 3585 & 619 &  125   	& 	$0.58_{-0.39}^{+0.32}$  & 	\cite{Knutson2011} \\
    GJ 9827 d \citep{Roy2023}               &   2.07    &   5.2  	&   1189   	&   6.2015  & 0.61  &   0.61 & 4269 & 621 &  142    &   $3.21_{-1.05}^{+0.99}$  &  	\cite{Rodriguez2018} \\
    TOI-674 b \citep{Brande2022}            &   5.25    &   23.6  	&  	838    	&   1.9771  & 0.42  &   0.42 & 3514 & 661 &  215  	&   $1.05_{-0.41}^{+0.46}$  & 	\cite{Murgas2021} \\
    HD 97658 b \citep{Guo2020}              &   2.12    &   8.3 	&  	1809  	&   9.4897 	& 0.73  &   0.85 & 5212 & 681 &  103    & 	$0.06_{-0.05}^{+0.74}$  &   \cite{Ellis2021} \\
    HAT-P-11 b \citep{Fraine2014}           &   4.73    &   25.74   &   1133 	&   4.8878 	& 0.75  &   0.81 & 4780 & 796 &  193    & 	$2.69_{-0.64}^{+0.59}$  &   \cite{Bakos2010} \\
    HD 106315 c \citep{Kreidberg2022}       &   4.35    &   15.2 	&   787   	&  21.0570 	& 1.30  & 	1.09 & 6327 & 811 &  281 	& 	$1.91_{-0.87}^{+1.14}$  &  	\cite{Barros2017} \\
    HIP 41378 b (This Work)                 &   2.9     &   8.29\tablenotemark{$\ddag$} 	&   965   	&  15.5712 	& 1.40  &   1.15 & 6199 & 904 &  255   	& 	$2.64_{-2.50}^{+2.62}$  & 	\cite{Vanderburg2016} \\
    HAT-P-26 b \citep{Wakeford2017}         &   6.3     &   18.75   &  	458 	&   4.2345  & 0.79 	& 	0.82 & 5079 & 909 &  541  	&  	$2.97_{-0.45}^{+0.40}$  &	\cite{Hartman2011} \\
    \enddata
    \label{tab:sample}
\caption{Planets in our sample and their relevant parameters. Scale heights were calculated assuming $\mu=3.05$ amu, and equilibrium temperatures calculated assuming an albedo of 0.3.}
\tablenotetext{*}{Value calculated as part of this analysis.}
\tablenotetext{\dag}{\cite{Knutson2014}}
\tablenotetext{\ddag}{Calculated using the \cite{Chen2017} mass-radius relationship}
\end{deluxetable*}

\section{Analysis}
\subsection{Planetary Parameters and Scale Heights}
From literature values for planetary and system parameters ($R_p$, $M_p$, $P_{orb}$, $M_*$), we recalculated the orbital semimajor axis $a$ (according to Eq.~\ref{eq:a}), planet surface gravity $g$ (according to Eq.~\ref{eq:surfg}), as well as planet equilibrium temperature $T_{eq}$ (according to Eq.~\ref{eq:teq}) for all planets in our sample, assuming a planetary albedo of $A=0.3$. 
\begin{equation}
    a = \left( P_{orb}^2 \frac{GM_*}{4\pi^2} \right)^{1/3}
    \label{eq:a}
\end{equation}

\begin{equation}
    g = \frac{GM_p}{R_p^2}
    \label{eq:surfg}
\end{equation}

\begin{equation}
    T_{eq} = \mathrm{T_{eff}} \sqrt{\frac{R_*}{2a}}(1 - A)^{1/4}
    \label{eq:teq}
\end{equation}

To calculate the scale heights of the atmospheres of these planets, we assumed $\mu = 3.05$ amu (as these Neptune-sized worlds are more likely to be nearer $100\times$ Solar metallicity than $1\times$ Solar, \cite{Welbanks2019}), and calculated $H$ according to Eq. \ref{eq:H}. 
Our calculated values are shown in Table \ref{tab:sample}.


\subsection{Atmospheric Model Fitting and Retrievals}



We used the open-source code \texttt{petitRADTRANS} \citep{Molliere2019} to conduct isothermal, free-chemistry atmospheric retrievals of the archival \hst{}/WFC3 spectra of the planets in our sample. Our retrieval framework fits for five parameters: planet mass, planet radius, isothermal atmospheric temperature, \water{} abundance ($\log_{10}X$, where $X$ is the \water{} mass fraction), and gray cloud deck pressure ($\log_{10}\mathrm{P}$, in bars). We placed Gaussian priors on planetary mass and radius based on the literature values and our calculated gravity distributions, a uniform prior on isothermal atmospheric temperature of $T_{eq}~\pm~200$~K, and placed uniform priors on the $\log_{10}$ cloud deck pressure (in bars, $\mathcal{U}(-6, 2)$) and $\log_{10}$ water abundance (as mass fractions, $\mathcal{U}(-6, 0)$) Our retrievals also included Rayleigh scattering from H$_2$ and He, as well as collisionally-induced-absorption of H$_2$-H$_2$ and H$_2$-He as continuum opacities. We fixed the reference pressure of our retrievals at 0.01 bar, and calculated model spectra at $R=1000$. 


For each planet, to measure \ah{}, we sampled 100 random model spectra from our retrievals and rebinned them to a resolution of $R=250$, using the flux-conserving resampling method in \texttt{astropy}'s \citep{AstropyCollaboration2013, AstropyCollaboration2018, AstropyCollaboration2022} \texttt{specutils} package. For each sampled spectrum $i$, we calculated $A_{H_i}~=~n_i~=~(\delta_{1.4,i}~-~\delta_{1.25,i})(R_*^2/2H_iR_{p,i})$, as re-arranged from Equation \ref{eq:delta}, where $\delta_{\lambda,i}$ is the transit depth of the sampled spectrum $i$ at wavelength $\lambda$, $R_*$ the stellar radius for the system in question, $R_{p,i}$ the retrieved radius of the planet for sampled spectrum $i$, $H_i$ the atmospheric scale height (calculated with the retrieved mass and radius values for each sampled spectrum), and $n_i=A_{H_i}$ the amplitude of the feature in numbers of scale heights $H_i$. The final \ah{} value and its confidence interval for a particular planet was calculated from the $16^{th}$, $50^{th}$, and $84^{th}$ percentiles of the sampled values for $n_i$.

\begin{figure*}
    \centering
    \includegraphics[]{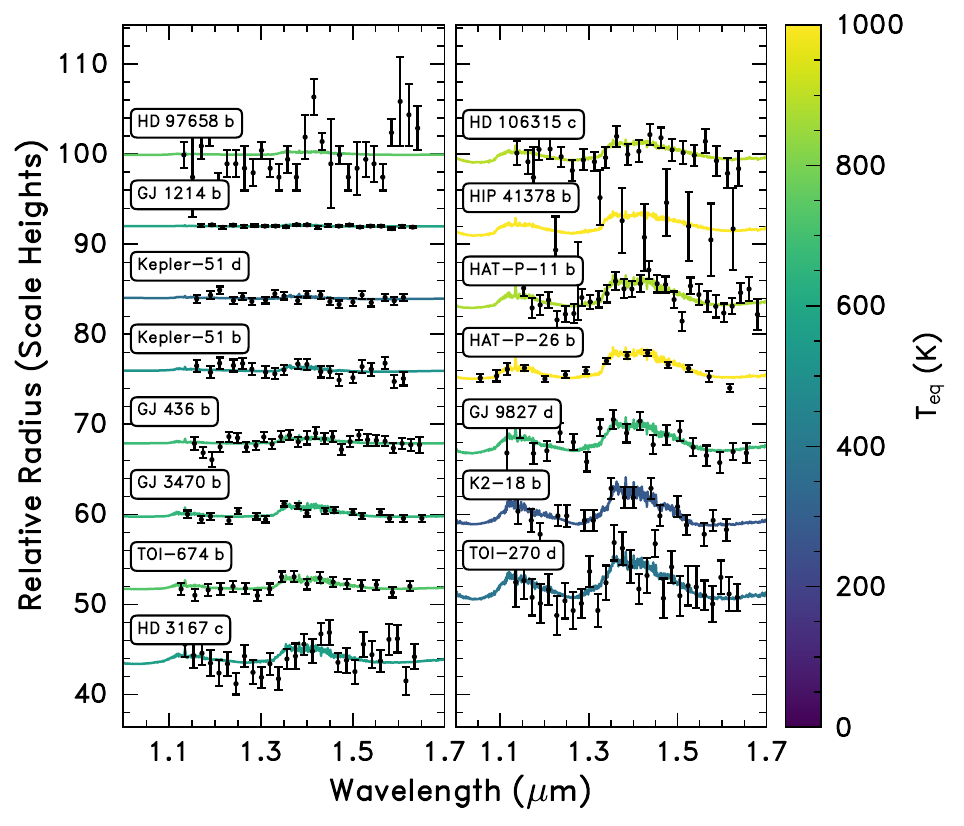}
    \caption{All \hst{}/WFC3 G141 transmission spectra of our exo-Neptunes, scaled to show the relative \ah{} values. The retrieved models are colored by the planetary equilibrium temperature, and the spectra are ordered by \ah{}.}
    \label{fig:spectra_stack}
\end{figure*}

The observed transmission spectra and atmospheric models for our planets are shown in Figure \ref{fig:spectra_stack}.

\subsection{Bayesian Trend Modeling}
We searched for trends relating \ah{} to various planetary parameters in our sample, such as equilibrium temperature, planet surface gravity, and stellar metallicity, but found no hints of useful relationships between any other than \ah{} and equilibrium temperature. We then worked to identify any possible temperature dependent trend in a hierarchical Bayesian framework, in order to both distinguish between possible polynomial models as well as quantify the inherent scatter in the data. We also exclude two planets in our sample, Kepler-51 b and d, from the trend fitting, as these so-called ``super-puffs'' are likely not in the same physical category as the rest of the sample due to their anomalously low densities and surface gravities. It's unclear whether these, along with GJ 1214 b and its exceptionally flat spectrum, are part of the same population as the other cold planets that have strong absorption features. Perhaps below a certain $T_{eq}$, the clarity of sub-Neptune atmospheres is essentially random with respect to $T_{eq}$, or perhaps something splits these planets into clear and cloudy populations. Studies of GJ 1214 b's atmosphere with \jwst{} continue to indicate it too may be unlike its peers, with especially high ($>1000\times$ Solar) metallicity \citep{Gao2023}. As such, we also exclude it from the final trend fitting. 

We aim to distinguish between several trend models: one with zero correlation between \ah{} and temperature (\ah{}~=~some~constant~$c$), a linear model (\ah{}~$=~bt+c$) and one with a parabolic trend (\ah{}~$=~at^2 + bt + c$), where $t$ is the equilibrium temperature. We also included an extra parameter $\sigma_i$ to account for unknown scatter (whether a result of stochastic astrophysical processes or various modeling insufficiencies in our polynomial trends) within our sample, added in quadrature to the measured \ah{} uncertainties as $\sigma_{\mathrm{fit}} = \sqrt{\sigma_{A_H}^2 + \sigma_i^2}$. 

We used broad uniform priors on our model parameters, and used the No U-Turn Sampler available in \texttt{exoplanet} \citep{Foreman-Mackey2021} for model fitting and posterior sampling, running 4 chains for 3000 tuning steps and 2000 draws each. After all sampler runs converged (as evidenced by the Gelman-Rubin statistic $\hat{R}$ reaching a value of 1), we calculated $\chi^2_\nu$ and BIC values for each model and found that the parabolic model

$$A_H = 2.3\times10^{-5} t^2 -0.028t + 9.5$$ was preferred to either the constant or linear models, with $\chi_\nu^2 = 2.0$ and BIC=$25.1$. The best-fit parabolic model also includes 0.56 scale height of added scatter. Our priors, fitted parameters, and model comparisons are shown in Table \ref{tab:model}. Including the Kepler-51 planets and GJ 1214 b results in a worse fit to the data, with $\chi^2_\nu=23$ and BIC=287, as well as increases the added scatter to 0.95 scale height.

\begin{table*}[]
    \centering
    \begin{tabular}{llcc}
        \hline \hline
         Model & Parameter &  Prior & Fitted Value \\ \hline
        \textbf{Constant}, $\chi^2_\nu = 5.9$, BIC$=67.9$ \\ \hline
         & $c$ & $\mathcal{U}(-50, 50)$ & $1.8^{+0.29}_{-0.27}$\\
         & $\sigma_i$ & $\mathcal{U}(0, 10)$ & $1.1^{+0.24}_{-0.20}$ \\  \hline
        \textbf{Linear}, $\chi^2_\nu = 6.8$, BIC$=73.5$ \\ \hline
         & $b$ & $\mathcal{U}(-50, 50)$ & $-7.1^{+15}_{-15} \times 10^{-3}$\\
         & $c$ & $\mathcal{U}(-50, 50)$ & $2.3^{+1.0}_{-1.0}$ \\
         & $\sigma_i$ & $\mathcal{U}(0, 10)$ & $1.1^{+0.25}_{-0.20}$ \\  \hline
        \textbf{Parabola}, $\chi^2_\nu = 2.0$, BIC$=25.1$ \\ \hline
         & $a$ & $\mathcal{U}(-50, 50)$ & $2.3^{+0.45}_{-0.47} \times 10^{-5}$\\
         & $b$ & $\mathcal{U}(-50, 50)$ & $-0.028^{+0.0058}_{-0.0055}$ \\
         & $c$ & $\mathcal{U}(-50, 50)$ & $9.5^{+1.7}_{-1.8}$ \\
         & $\sigma_i$ & $\mathcal{U}(0, 10)$ & $0.56^{+0.20}_{-0.15}$ \\  \hline
    \end{tabular}
    \caption{Bayesian model fitting priors, fitted parameters, and model comparisons.}
    \label{tab:model}
\end{table*}


\section{Model Comparisons}

\begin{figure*}[htb!]
    \centering
    \includegraphics[width=0.9\textwidth]{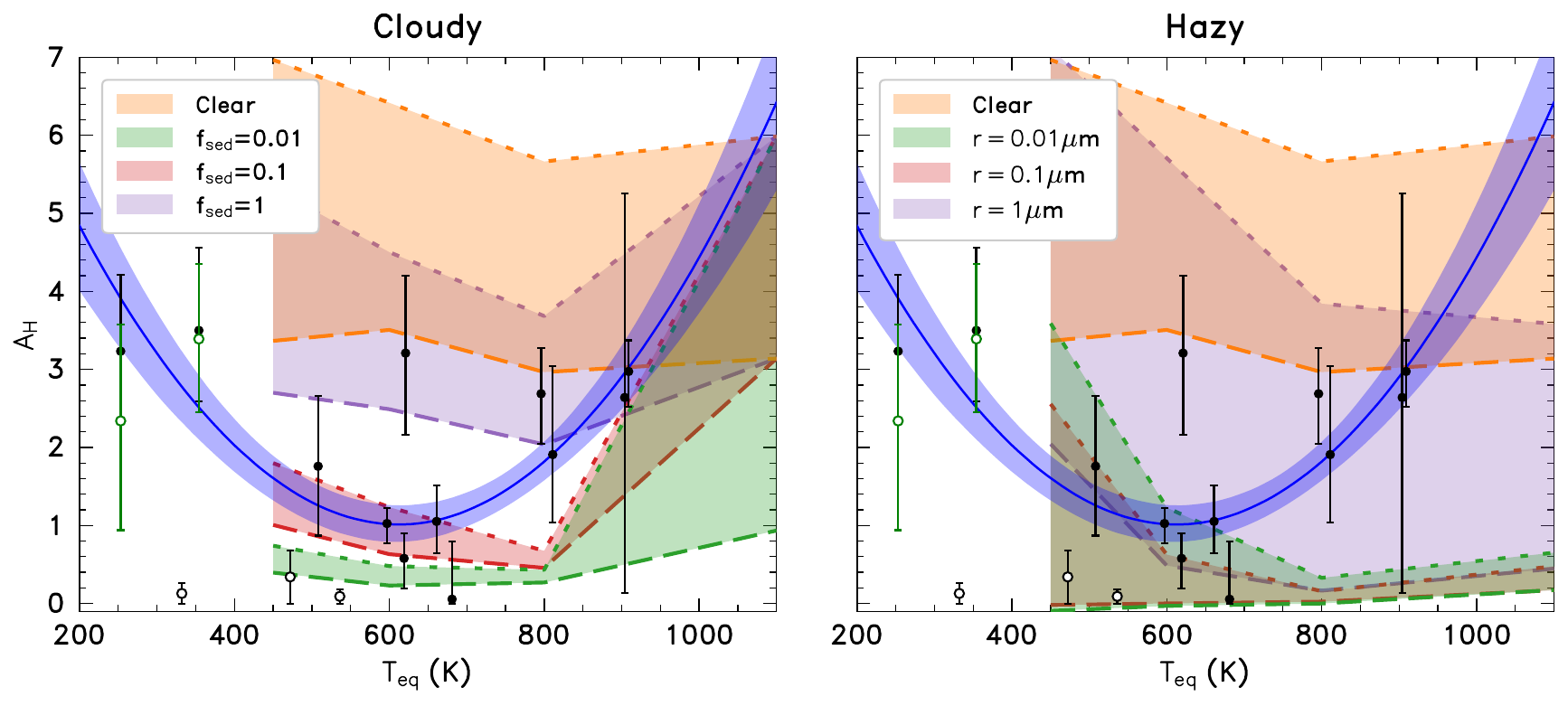}
    \caption{Retrieved spectral feature amplitude, \ah{}, compared to clear, cloudy, and hazy atmosphere models from \cite{Morley2015}. For the clear and cloudy models, the dotted lines indicate a $100\times$ Solar metallicity atmosphere, and the dashed lines a $300\times$ Solar atmosphere. All hazy models were calculated at $50\times$ Solar metallicity, the dotted lines are 1\% haze precursor conversion, and the dashed lines are 10\%. The green open markers, shown for context, are the measured \methane{}-retrieved \ah{} values for K2-18 b and TOI-270 d. The black open markers, shown for context, are the Kepler-51 planets and GJ~1214~b, which are excluded from the trend analysis.}
    \label{fig:clear_compare}
\end{figure*}

After identifying the best-fit parabolic trend, we compared our findings to a grid of sub-Neptune atmospheric models from \cite{Morley2015}. This grid modeled the atmosphere of GJ~1214~b in clear, cloudy, and hazy configurations, across various metallicities (from $50\times$ to $1000\times$ Solar) and cloud/haze parameters. The grid was also calculated at varying insolations (compared to GJ 1214 b's actual insolation), corresponding to $\sim450$~K at the low end to 1100~K on the high end. As with our retrievals, we resampled the model spectra to $R=250$, and calculated \ah{} using the same GJ 1214 b planetary parameters (but recalculating $H$ according to the model $T_{eq}$). The equilibrium cloud models include salt and sulfide species (ZnS, KCl, and Na$_2$S) calculated at varying f$_{\mathrm{sed}}$ and metallicity, and are more fully described in Section 2.3 of \cite{Morley2015}. The haze models were all calculated at $50\times$ Solar metallicity, included soots with five precursor species (C$_2$H$_2$, C$_2$H$_4$, C$_2$H$_6$, C$_4$H$_2$, and HCN), calculated at varying particle sizes and precursor conversion fractions, and are more fully described in Section 2.4 of \cite{Morley2015}. 
Using the $100\times$ and $300\times$ Solar cloudy models and the 1\% and 10\% haze precursor conversion fractions to bracket reasonable parameter ranges for our sample, we clearly see in Figure \ref{fig:clear_compare} that almost all of our sample planet spectra strongly attenuated by atmospheric aerosols compared to the clear atmosphere models.



\section{Results}

We calculate $\chi^2_\nu$ values for our planets with $T_{eq}~>~450$~K compared to the various contours for our clear, cloudy, and hazy models to find which set of models is most consistent with our observed \ah{} values. These model comparisons are shown in Table \ref{tab:model_comps}. We find that all our planets are likely to be attenuated by atmospheric aerosols (sometimes very significantly, up to $\sim6$ scale heights). This attenuation is strongest between $\sim500$ and $\sim700$~K, corresponding to a range of especially efficient aerosol production identified by theoretical models \citep{Morley2015, Gao2020}. The attenuation is also stronger for the $100\times$ Solar models compared to the $300\times$ Solar models. 

Even so, we note some interesting points. First, with respect to the cloudy models, the majority of our planets have low sedimentation efficiency, f$_{\mathrm{sed}}=0.1$ or lower, and the entire sample is most consistent with $\mathrm{f_{sed}} = 0.01$ and $100\times$ Solar metallicity. With respect to the hazy models, our sample is most consistent with small haze particles ($r=0.01\mu$m) and small (1\%) precursor conversion rates. Overall, the cloudy models fit better than the hazy models, although the sparsity of the grid means this is a relatively low-significance finding. More detailed modeling efforts may clarify this, as discussed in Section \ref{sec:priorities}. 
\begin{table}[]
    \centering
    \begin{tabular}{lr}
        \hline \hline
         Model & $\chi^2_\nu$ \\ \hline
        \textbf{Clear Models} \\ \hline
        $100\times$ Solar & 149\\
        $300\times$ Solar &  32\\ \hline
        \textbf{Cloudy Models} \\ \hline
        $\mathrm{f_{sed}} = 1,~100\times$ Solar & 62\\
        $\mathrm{f_{sed}} = 1,~300\times$ Solar & 13\\ \hline
        $\mathrm{f_{sed}} = 0.1,~100\times$ Solar & 2.9\\
        $\mathrm{f_{sed}} = 0.1,~300\times$ Solar & 4.7\\ \hline
        $\mathrm{f_{sed}} = 0.01,~100\times$ Solar & 3.8\\
        $\mathrm{f_{sed}} = 0.01,~300\times$ Solar & 9.0\\ \hline 
        \textbf{Hazy Models} \\ \hline
        $\mathrm{r} = 1\mu$m, 1\% conversion & 102\\
        $\mathrm{r} = 1\mu$m, 10\% conversion & 8.6\\ \hline
        $\mathrm{r} = 0.1\mu$m, 1\% conversion & 8.1\\
        $\mathrm{r} = 0.1\mu$m, 10\% conversion & 12.6\\ \hline
        $\mathrm{r} = 0.01\mu$m, 1\% conversion & 7.4\\
        $\mathrm{r} = 0.01\mu$m, 10\% conversion & 12.9\\ \hline  \hline
    \end{tabular}
    \caption{Atmospheric models and model comparisons.}
    \label{tab:model_comps}
\end{table}

Second, while they were not included in the trend fitting, the Kepler-51 planets cannot be totally ignored. While these ``super-puffs'' clearly stand out as exceptionally low-density planets, other temperate, puffy sub-Neptunes have been discovered since (e.g. TOI-700~c, WASP-193~b), which will also be amenable to atmospheric transmission spectroscopy. Observing as many cold planets as possible is sorely needed to determine whether our observed trend holds. These observations would identify whether cold sub-Neptunes exist in a continuum between the slightly attenuated atmospheres which follow our trend and the especially cloudy and hazy planets like the Kepler-51 planets and GJ 1214 b. If these are outliers among the broader population, it bodes well for future studies of habitable zone systems. Perhaps there is a range of temperatures where cloud rainout is efficient (especially for heavier species like ZnS and KCl) while low insolation means haze production is not, giving especially clear views into these temperate planets' atmospheres. 

Finally, we find that, among our sample of Neptune-sized exoplanets, there is an intrinsic scatter in the clarity of these atmospheres of $\sim 0.5$ scale height. This scatter may be evidence that our trend model choices are insufficient, or possibly be due to some randomness in the outcomes of planet formation and evolutionary processes. 
Perhaps the trend identified is analogous to a ``main sequence'' of planetary atmospheres, and (taking into account the super-puffs) future observations with \jwst{} will be able to confirm or reject our identification of this scatter.

We also note that, while we used water vapor as the major opacity source in this analysis, methane absorption has significant overlap with water in near-IR bands, and may dominate compared to water vapor in certain exoplanetary atmospheres $<600$~K \citep{Bezard2022}. In fact, this has been identified in K2-18~b, previously thought to have water vapor in its atmosphere from \hst{} data \citep{Benneke2019a}, but now shown instead to have methane from broad-wavelength \jwst{} observations \citep{Madhusudhan2023}. With this in mind, we also ran methane-only retrievals for K2-18~b and TOI-270~d to compare to our water-only retrievals and found the methane-calculated \ah{} values to be consistent within the $1\sigma$ uncertainties. The substitution of \methane{} for \water{} in these coolest planets does not change our conclusions. These are shown in Figure~\ref{fig:clear_compare} as open green markers.

These results echo previous analyses: our model comparisons provide evidence for a cloud-based mechanism shaping this trend (predicted by \cite{Fu2017}) over a haze-driven trend (predicted by \cite{Crossfield2017} and \cite{Yu2021}). Qualitatively, our quadratic trend behaves similarly to previous second-order trends fit to samples that extend to low temperatures \citep{Yu2021, Edwards2023}, showing significant increases in atmospheric clarity at cold ($<500$K) and hotter ($>700$K). The major differences between these trends are primarily due to sampling, as the inclusion of particular planets (like Kepler-51 d in \cite{Yu2021} and \cite{Edwards2023}) can shift the parabola minimum to colder equilibrium temperatures. Robustness against planetary samples is encouraging, as it appears the observational predictions made by these trends will be useful for a broad range of future spectroscopic studies. Most of our measured $A_H$ values are also consistent to previous analyses within error, although our method differs to previous analyses due to our retrieval framework. This confirms that, as an observational metric, $A_H$ is robust against methodological changes (even if the physical insights gained about any particular planet transmission spectrum from $A_H$ alone appear to be limited).

\section{Community Priorities}
\label{sec:priorities}

Despite the proliferation of \hst{} WFC3 transmission spectra in recent years, significant trends relating the clarity of these atmospheres to planetary and system parameters still evade us. WFC3, while especially sensitive to water vapor and uniform cloud absorption within a narrow spectral range, cannot characterize the full panoply of absorbers present in exoplanet atmospheres (perhaps most significantly shown by K2-18~b's \jwst{} methane detection in \cite{Madhusudhan2023} over the \hst{} water detection from \citep{Benneke2019a}), nor precisely constrain physical cloud models. As already shown from the first year of \jwst{} analyses, exoplanetary atmospheres are complex, dynamic, and contain clear evidence of atmospheric physics beyond anything \hst{} could provide \citep{JWSTTransitingExoplanetCommunityEarlyReleaseScienceTeam2023, Rustamkulov2023, Alderson2023, Feinstein2023, Ahrer2023, Tsai2023}. 

Among the scheduled and completed \jwst{} programs (through Cycle 2), there are 106 unique exoplanet systems. Of the planets in these systems, 36 fit our sample radius criteria ($\mathrm{R_\mathrm{pl}}\in[2,6]$~$\mathrm{R_\oplus}$). 8 of the planets we analyzed in this work are on this list as well, although these alone are not sufficient to adequately sample the range of equilibrium temperatures. We need to add more exoplanets to the \jwst{} observational sample, especially temperate planets that are large enough to sustain extended atmospheres, so we can confirm the~$<500$K leg of the trend observed in this work.

We also need significantly improved model grids for sub-Neptune atmospheres. Aerosol production is a complex process, and low-resolution observations have historically been unable to reveal much about these. New \jwst{} observations, with higher resolution and much broader wavelength range, have already identified signatures of wavelength-dependent silicate cloud absorption in the planetary-mass object VHS~1256~b \citep{Miles2023}. Haze precursors like SO$_2$ have been identified in the Hot Jupiter WASP-39 b \citep{Rustamkulov2023, Alderson2023, Tsai2023}, and GJ~1214~b's high-altitude aerosols have been found to be very reflective \citep[$A=0.51$,][]{Kempton2023}. Extending modeling efforts to colder planets, across a wide range of metallicities, cloud species, particle sizes, atmospheric mixing, and other parameters should enable us to constrain these processes in these planets' atmospheres.   

\section{Conclusions}

We have identified a coherent trend relating the clarity of Neptune-sized planetary atmospheres to their equilibrium temperatures, as well as modeling the intrinsic scatter in planetary clarity measurements, across an $\sim800$K temperature range. These findings show behavior similar to previous analyses, again finding a parabolic trend with clearer atmospheres at cooler ($<500$K) and hotter ($>700$K), with a minimum between $500-700$ K. We reproduce this behavior for our restricted sample of Neptune-like planets compared to more diverse samples in previous analyses, implying the observational implications of this work should be applicable to many future \hst{} and \jwst{} targets. Compared to models, all observed planetary atmospheres are consistent with attenuation by at least some cloud cover as a result of vertically extended clouds or hazes with small particles ($\mathrm{f_{sed}} = 0.1$, $r=0.01\mu$m), and have relatively high metallicity ($\sim100\times$ solar). Confirmation or rejection of cold-temperature theoretical predictions still eludes us, as we need updated self-consistent modeling efforts across a range of atmospheric parameters for these cold planets. \jwst{} observations in the next few cycles should both continue to test our identified trend, and help identify the atmospheric diversity that may be present among cold ($T_{eq} < 500$~K) planets.

\section*{Acknowledgments}

We thank Luis Welbanks for helpful discussions and the anonymous referee whose comments significantly improved the work.

\appendix
\section{HIP 41378 b}

HIP 41378 b was included as a target in \hst{} program GO-15333, which also observed HD 3167 c, GJ 9827 d, TOI-674 b, and HD 106315 c. 

We observed two transits of HIP~41378~b with \hst{}'s WFC3 instrument, using the G141 grism, before dropping this planet in favor of TOI-674 b \citep{Brande2022} due to scheduling difficulties. We followed standard data reduction procedures for \hst{}/WFC3 time series data, as documented in \cite{Kreidberg2022}. Following common practice, we fit the light curves simultaneously with a transit model and an instrument systematics model. Our transit model included variable $R_p/R_*$, and other system parameters were fixed to the best-fit values from \citep{Berardo2019}. The limb darkening parameters were fixed on predictions from a PHOENIX stellar model at the effective temperature of the host star. The instrument systematics model consisted of a visit-long linear slope, a scaling factor between upstream and downstream spatial scans, and exponential ramps fit to each orbit \citep[following the model ramp parameterisation from][]{Kreidberg2022}. The observed transmission spectrum is given in Table \ref{tab:hip41378b}. The raw data for HIP 41378 b is available at Mikulski Archive for Space Telescopes (MAST) at the Space Telescope Science Institute at the following DOI:\dataset[10.17909/ryg6-3c46]{https://doi.org/10.17909/ryg6-3c46}.

\begin{table}[h!]
    \centering
    \begin{tabular}{cc}
        \hline \hline
        Wavelength  & Transit Depth $(R_p/R_*)^2$ \\
        ($\mu$m) & (ppm) \\ \hline
        1.175 & $347 \pm 32.7$ \\
        1.225 & $273 \pm 32.2$ \\
        1.275 & $244 \pm 35.5$ \\
        1.325 & $326 \pm 28.6$ \\
        1.375 & $302 \pm 32.7$ \\
        1.425 & $285 \pm 33.2$ \\
        1.475 & $320 \pm 34.9$ \\
        1.525 & $296 \pm 34.7$ \\
        1.575 & $283 \pm 35.6$ \\
        1.625 & $294 \pm 41.1$ \\ \hline
    \end{tabular}
    \caption{HIP 41378 b \hst{}/WFC3 G141 Transmission Spectrum}
    \label{tab:hip41378b}
\end{table}

%

\vspace{5mm}
\facilities{NASA Exoplanet Archive \citep{ExoArchive}}


\software{Astropy: \citep{AstropyCollaboration2013, AstropyCollaboration2018, AstropyCollaboration2022}, \texttt{exoplanet}: \citep{Foreman-Mackey2021}}

\bibliography{bibliography}{}
\bibliographystyle{aasjournal}



\end{document}